\documentclass[a4paper,11pt]{article}

\usepackage{graphicx}
\usepackage{url}
\usepackage[utf8]{inputenc} % allow utf-8 input
\usepackage[T1]{fontenc}    % use 8-bit T1 fonts
\usepackage{hyperref}       % hyperlinks
\usepackage{url}            % simple URL typesetting
\usepackage{multirow}

\usepackage{soul} % strikethrough \st{Hello world}

\usepackage[numbers]{natbib}

\usepackage{float}  % To place the figures where we want
\usepackage[normalem]{ulem}

\usepackage{xspace}

\input{glyphtounicode}  % With this ``fi'' and other ligatures are copied correctly from the PDF
\usepackage{color}
\usepackage{colortbl}
\usepackage{xspace}
\definecolor{lightgrey}{gray}{0.35}
\definecolor{orange}{rgb}{1,0.8,0.5}
\definecolor{lightorange}{rgb}{1,0.9,0.5}
\definecolor{otherorange}{rgb}{1,0.8,0.6}

\definecolor{brown}{RGB}{150,70,0}
\definecolor{green}{RGB}{127,255,127}
\definecolor{darkgreen}{RGB}{0,127,0}
\definecolor{blue}{RGB}{127,127,255}
\definecolor{lightblue}{RGB}{150,150,255}
\definecolor{darkblue}{RGB}{40,40,177}
\definecolor{red}{RGB}{255,90,90}
\definecolor{violet}{RGB}{150,70,130}
\definecolor{darkred}{RGB}{200,50,50}
\definecolor{grey}{RGB}{127,127,127}
\definecolor{pink}{RGB}{255,180,180}

\newcommand{\hide}[1]{}
\newcommand{\cut}[1]{}

\newcommand{\Exploitation}{Exploitation} %Interpretation and Use, Interpretation and Deployment

\ifdefined\NOREFERENCES
\renewcommand{\cite}[1]{}
\renewcommand{\citep}[1]{}
\fi

\newcommand*{\NOTES}{}

\ifdefined\NOTES
    
    \newcommand{\insidenote}[1]{\textcolor{darkgreen}{[\small{#1}] \marginpar{$\bullet$}}}
    \renewcommand{\insidenote}[1]{\textcolor{darkgreen}{[\small{#1}]}}
    
    \newcommand{\sidenote}[1]{\textcolor{darkgreen}{${\leftarrow}\hspace{0pt}{\bullet}$}\marginpar{\textcolor{darkgreen}{${\leftarrow}\hspace{0pt}{\bullet}$}\\\tiny{\textcolor{darkgreen}{#1}}}}

    \newcommand{\sidenoteJose}[1]{\textcolor{violet}{${\leftarrow}\hspace{0pt}{\bullet}$}\marginpar{\textcolor{violet}{${\leftarrow}\hspace{0pt}{\bullet}$}\\\tiny{\textcolor{violet}{Jose: #1}}}}

    \newcommand{\sidenotePadhraic}[1]{\textcolor{blue}{${\leftarrow}\hspace{0pt}{\bullet}$}\marginpar{\textcolor{blue}{${\leftarrow}\hspace{0pt}{\bullet}$}\\\tiny{\textcolor{blue}{Padhraic: #1}}}}

    \newcommand{\sidenoteHolger}[1]{\textcolor{blue}{${\leftarrow}\hspace{0pt}{\bullet}$}\marginpar{\textcolor{blue}{${\leftarrow}\hspace{0pt}{\bullet}$}\\\tiny{\textcolor{blue}{Holger: #1}}}}

    \newcommand{\sidenoteChris}[1]{\textcolor{blue}{${\leftarrow}\hspace{0pt}{\bullet}$}\marginpar{\textcolor{blue}{${\leftarrow}\hspace{0pt}{\bullet}$}\\\tiny{\textcolor{blue}{Chris: #1}}}}

    \newcommand{\sidenoteLuc}[1]{\textcolor{blue}{${\leftarrow}\hspace{0pt}{\bullet}$}\marginpar{\textcolor{blue}{${\leftarrow}\hspace{0pt}{\bullet}$}\\\tiny{\textcolor{blue}{Luc: #1}}}}

    \newcommand{\sidenoteTijl}[1]{\textcolor{blue}{${\leftarrow}\hspace{0pt}{\bullet}$}\marginpar{\textcolor{blue}{${\leftarrow}\hspace{0pt}{\bullet}$}\\\tiny{\textcolor{blue}{Tijl: #1}}}}

    \newcommand{\notes}[1]{\noindent\textit{#1}}

\else
    \newcommand{\insidenote}[1]{}
    \newcommand{\sidenote}[1]{}
    \newcommand{\sidenoteJose}[1]{}
    \newcommand{\sidenotePadhraic}[1]{}
   \newcommand{\sidenoteHolger}[1]{}
    \newcommand{\sidenoteChris}[1]{}
    \newcommand{\sidenoteLuc}[1]{}
    \newcommand{\sidenoteTijl}[1]{}
    \newcommand{\notes}[1]{}

\fi

\usepackage{framed}

\newcommand{\quadrant}{quadrant\xspace}
\newcommand{\quadrants}{quadrants\xspace}

\long\def\comment#1{}

\renewcommand{\sout}[1]{\unskip}

\begin{document}

\ifdefined\NOTITLE
\title{}
\else
\title{
\makebox[0pt][l]{\raisebox{1in}[0pt][0pt]{\hspace{-1.75in}
\parbox{6in}{\small Final m/s version of paper published in
  \emph{Communications of  the ACM} 65(3) pp 76-87 (2022). Please cite the
  journal version, which contains the final version of the figures.}}}
Automating Data Science:\\ Prospects and Challenges\\$\:$} \fi

\author{

\ifdefined\NOAUTHORS
\else
Tijl De Bie\\
\scriptsize{ IDLab -- Dept. of Electronics and Information Systems } \\
\scriptsize{Ghent University}\\ 
\scriptsize{Belgium}\\
\hspace*{4cm}
%\alignauthor
\and
Luc De Raedt\\
\scriptsize{Dept. of Computer Science}\hspace{1cm}\scriptsize{AASS}\\
%\scriptsize{KU Leuven Institute for AI}\\
\hspace*{1cm}\scriptsize{KU Leuven}\hspace{1,2cm}\scriptsize{\" Orebro University}\\ 
\hspace*{1cm}\scriptsize{Belgium}\hspace{2,3cm}\scriptsize{Sweden}\\ \\
\hspace*{4cm}
%\alignauthor
\and
José Hernández-Orallo\\
\scriptsize{vrAIn}\\
%Valencian Research Institute for Artificial Intelligence\\
\scriptsize{Universitat Polit\`ecnica de Val\`encia}\\ 
\scriptsize{Spain}\\
\hspace*{4cm}
%\alignauthor
\and
Holger H. Hoos\\
\scriptsize{LIACS}\\
\scriptsize{Universiteit Leiden}\\
\scriptsize{The Netherlands}\\
\hspace*{4cm}
%\alignauthor
\and
Padhraic Smyth\\
\scriptsize{Department of Computer Science}\\
\scriptsize{University of California, Irvine}\\
\scriptsize{USA}\\
\hspace*{4cm}
%\alignauthor
\and
Christopher K. I.  Williams\\
\hspace*{2mm}\scriptsize{School of Informatics}\hspace{1cm}Alan Turing Institute\\
\hspace*{-10mm}\scriptsize{University of Edinburgh}\hspace{15mm}London\\
\scriptsize{United Kingdom}\hspace{15mm}\scriptsize{United Kingdom}\\
\hspace*{4cm}
%\alignauthor
\fi
}

\ifdefined\NODATE
\date{}
\else
\date{\today}
\fi

\maketitle

%------------------------------------------------------------------

\vspace{2cm}

\ifdefined\NOFLOATS

\else

{ {\textcolor{black}{\em 
Given the complexity of typical data science projects and the associated demand for human expertise, automation has the potential to transform the data science process.  
}}}

\vspace{1cm}

%\vspace{0.2cm}
%\end{abstract}
% The classification Scheme, General Terms and Keywords are not appropriate for CACM so comment them out.

%------------------------------------------------------------------

\begin{framed}
\vspace{-0.1cm}
\noindent{%\large
 \bf   Key insights}
\begin{itemize}
    \item Automation in data science aims to  facilitate and transform the work of data scientists, not to replace them. 
    \item Important parts of data science are already 
being automated, especially in the modeling stages, where techniques such as automated machine learning (AutoML) are gaining traction.
  \item Other aspects are harder to automate, not only because of technological challenges, but because open-ended and context-dependent tasks require human interaction.
\end{itemize}
\vspace{-0.3cm}
\end{framed}
\fi

\newpage

\section*{Introduction}

Data science covers the full spectrum of deriving insight from data,
from initial data gathering and interpretation, via processing and
engineering of data, and exploration and modeling, to eventually producing
novel insights and decision support systems. 

Data science can 
be viewed as overlapping or broader in scope than other  data-analytic methodological disciplines, such as statistics, machine learning, databases, or visualization \cite{donoho201750}.

To illustrate the breadth of data science, consider, for example, the problem of 
recommending items (movies, books or other products) to customers.  
While the core of these applications 
can consist of algorithmic techniques such as matrix factorization, 
a deployed system will involve a much wider range of technological and human considerations. 
These range from scalable back-end transaction systems that retrieve customer and product data in real time,  experimental design for evaluating system changes, causal analysis for understanding the effect of interventions, to the human factors and psychology that underlie how customers react to visual information displays and make decisions. 

As another example, 
in areas such as astronomy, particle physics, and climate science, there is a rich tradition of building computational pipelines to support data-driven discovery and hypothesis testing. For instance, geoscientists use monthly global landcover maps based on satellite imagery at 
sub-kilometer  
resolutions to better understand 
how the earth's surface is changing over time \cite{wulder2018land}.
 These maps are interactive and browsable, and they are the result of a complex data-processing pipeline, in which terabytes to petabytes of raw sensor and image data are transformed into databases of automatically detected and annotated objects and information. This type of pipeline involves many steps, in which 
human decisions and insight are critical, such as instrument calibration, removal of outliers, and classification of pixels.

The breadth and complexity of these and many other  
data science scenarios 
means that the modern data scientist requires broad knowledge and experience across a multitude of topics. 
Together with an increasing demand for data analysis skills, this has led to a shortage of trained data scientists with appropriate background and experience, and 
significant market competition for limited expertise.  Considering this bottleneck, it is not surprising that there is increasing interest in 
automating parts, if not 
all, of the data science process. 
This desire and potential for automation is the focus of this article.

As illustrated in the examples above, data science is 
a complex process, driven by the character of the data being analyzed 
and by the questions being
asked, and is often  highly exploratory and iterative in nature.
Domain context can play a key role in these exploratory steps, even in relatively well-defined processes such as predictive modeling (e.g., as characterized by CRISP-DM \cite{chapman2000crisp}) where, for example, human expertise in defining relevant predictor variables can be critical.

\ifdefined\NOFLOATS
\else
\begin{figure}%[ht]
\begin{center}
\includegraphics[width=0.75\textwidth]{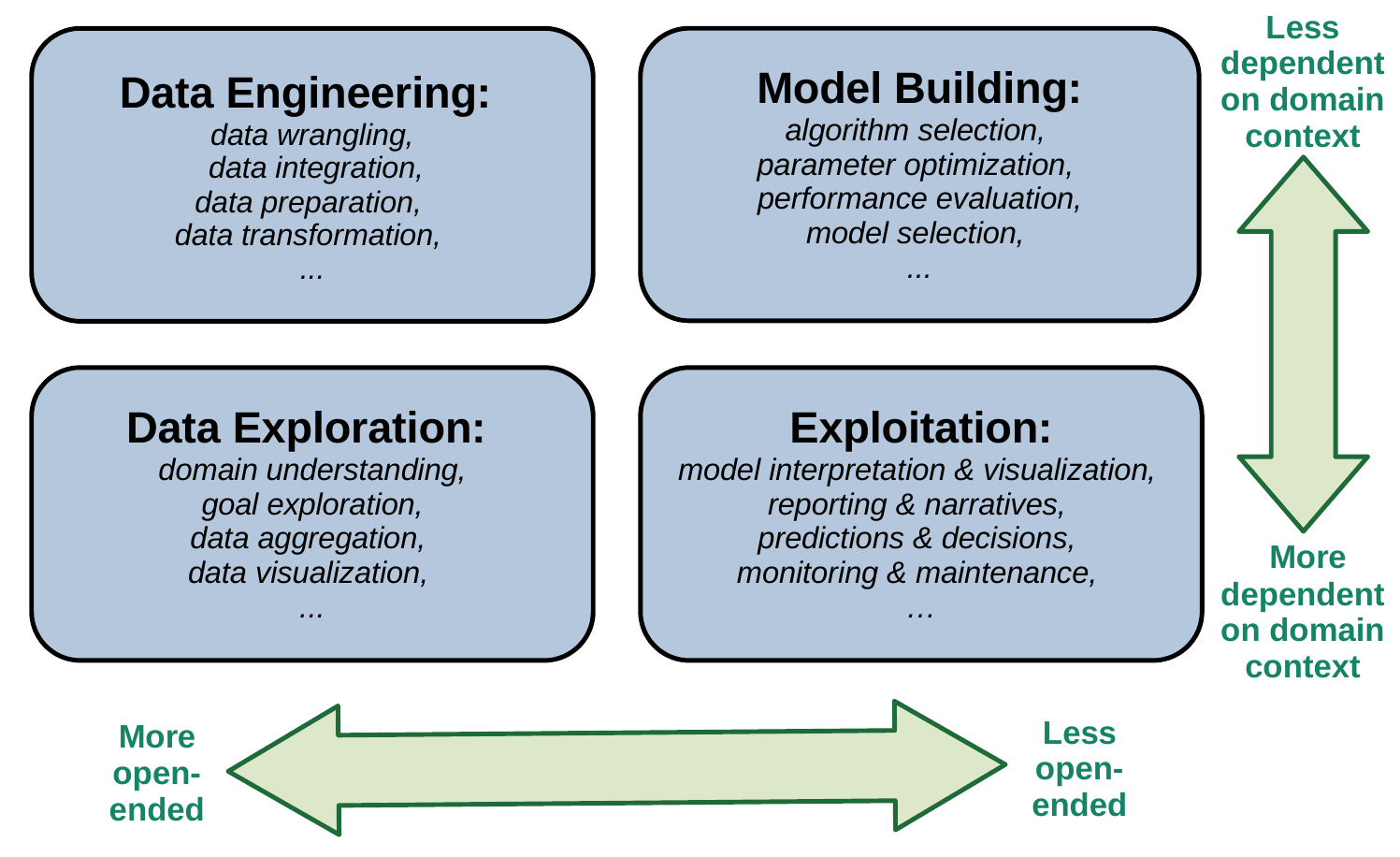} \hfill
\end{center}
 %\hspace{1cm}
 \vspace{-1cm}
\caption{The four data science \quadrants we use in this article to illustrate different areas where automation can take place. The vertical dimension
determines the degree of dependence on domain context, usually introduced through human interaction. 
The horizontal dimension determines the degree to which a process is open-ended. Some activities, such as  data augmentation and feature engineering, are situated in data engineering near  the boundary with data exploration.} 
\label{fig:quadrant}
\end{figure}
\fi

With this in mind, Figure \ref{fig:quadrant} 
 provides a conceptual framework to guide  
our discussion  of automation in data science, including aspects  that are already being automated as well as aspects that are potentially ready for automation. 
The vertical dimension of the figure reflects the degree to which domain \sout{knowledge or} context plays a role in the process. 
Domain context not only includes domain knowledge 
but also human factors, 
such as the 
interaction of humans with the technology \cite{amershi2019guidelines}, the side effects on users and non-users, and all the safety and ethical issues, including algorithmic bias. 
These factors 
have various effects on data understanding and the impact of the extracted knowledge, once deployed, and are often addressed or supervised with humans in the loop.

The lower \quadrants
of {\em Data Exploration} and {\em \Exploitation} are typically  closely coupled to the application domain, 
while the upper \quadrants of {\em Data Engineering} and {\em Model Building} are often  more domain-agnostic. 
The horizontal axis 
characterizes the degree to which 
different activities in the overall process range from being more
open-ended to more precisely specified, such as having well-defined 
goals, clear modeling tasks and  measurable performance indicators.  {\em Data Engineering} and {\em Data Exploration} are often not 
precisely specified and are quite iterative in nature,
while {\em Model Building} and {\em \Exploitation} are often defined
more narrowly and precisely. In classical goal-oriented projects, 
the process often consists of activities 
in the following order: Data Exploration, Data Engineering, Model Building and Exploitation. In practice, however, 
these trajectories can be much more diverse and exploratory, with practitioners navigating through activities in these quadrants in different orders and in an iterative fashion (see, e.g., \cite{Plumed2019}).

From the layout of Figure \ref{fig:quadrant}
we see, for example, that  {\em Model Building} is where we might
expect automation to have the most direct impact 
---which is indeed the case with the success of automated machine learning (AutoML). However, 
much of this impact has occurred for modeling approaches based on supervised learning,  and automation 
is still far less developed for other kinds of learning
or modeling tasks.

Continuing our discussion of Figure \ref{fig:quadrant},   {\em Data Engineering} tasks are estimated to often take 80\% of the human effort in a typical data analysis project \cite{dasu-johnson-03}. 
As a consequence it is natural to expect that automation could play a major role in reducing this  human effort. However, efforts to automate {\em Data Engineering} tasks  have had less success to date compared to efforts in automating {\em Model Building}.

{\em Data Exploration} 
involves identifying relevant questions given a dataset, interpreting the structure of the data, understanding the constraints provided by the domain as well as the data analyst's background and intentions, and identifying issues related to data ethics, privacy, and fairness. 
Background knowledge and human judgement are key to success. 
Consequently, it is not surprising that {\em Data Exploration} poses the greatest challenges for automation.  

Finally, {\em \Exploitation} turns actionable insights and predictions into decisions.  
As these may have a significant impact, some level of oversight and
human involvement is often essential; e.g., new AI techniques can
bring new opportunities in automating the reporting and explanation of
results \cite{lloyd2014a}, as discussed below. 

Broadly speaking, automation in the context of data science is more or less challenging depending on the form it takes, ranging in complexity depending on whether it involves a single task or an entire iterative process, or whether partial or complete automation is the goal. 
\begin{enumerate}
    \item A first form of automation---{\em mechanization}---occurs when a task is so well specified that there is no need for human involvement. 
Examples of such tasks include  running a clustering algorithm or standardizing the values in a table of data. 
This can be done by functions or modules in low-level languages, or as part of statistical and algorithmic packages that have traditionally been used in data science. 
\item A second form of automation---{\em composition}---deals with strategic sequencing of tasks or integration of different parts of a task. 
Support for code or workflow reuse is available in more sophisticated tools that have emerged in recent years,  from interactive workflow-oriented suites (such as KNIME, RapidMiner, 
IBM Modeler, SAS Enterprise Miner, Weka Knowledge Flows and Clowdflows) 
to 
high-level programming languages and environments commonly used for data analysis and model building (such as  R, Python, Stan, BUGS, TensorFlow and PyTorch). 
\item 
Finally, a third form of automation---{\em assistance}---derives from the production of elements such as visualizations, patterns, explanations, etc., that are specifically targeted at supporting human efficiency. This  includes a constant monitoring of what humans are doing during the data science process, so that an automated assistant can 
identify inappropriate choices,
make recommendations, and so on. 
While some limited form of assistance is already provided in interactive suites such as KNIME and RapidMiner, 
the  challenge is to extend this assistance to the entire data science process. 
\end{enumerate}

\noindent Below, we organize our discussion into sections corresponding to the four \quadrants from Figure \ref{fig:quadrant}, highlighting the three forms of automation where relevant.
Because the activities are arranged into quadrants rather than stages following a particular order,  
we begin with {\em Model Building}, which appears most amenable to automation, and then discuss 
the other \quadrants.

\section*{Model Building: The Success Story of AutoML} \label{sec:automl}

In the context of building models (see Figure 1), machine learning methods feature prominently in the toolbox of the data scientist, particularly because they tend to be formalized in terms of objective functions that directly relate to well-defined task categories.

Machine learning methods have become very prominent over the last two decades, including relatively complex methods, such as deep learning. 
Automation of these machine learning methods, which has given rise to a research area known as AutoML, is arguably the most successful and visible application to date of automation within the overall data science process (see, e.g., \cite{HutEtAl19}). 
It assumes, in many cases, that sufficient amounts of high-quality data are available; satisfying this assumption typically poses challenges, which we address in later sections of this article (see also \cite{RatEtAl16}).

While there are different categories of machine learning problems and
methods, including supervised, unsupervised, semi-supervised and
reinforcement learning, the definition of the target function and its
optimization is most straightforward for supervised learning (as illustrated in the box  ``From Machine Learning to Automated Machine Learning''). 
Focusing on supervised learning, 
there are many methods for accomplishing this task,  
often with multiple hyperparameters, whose values can have substantial impact on the prediction accuracy of a given model.

\ifdefined\NOFLOATS
\begin{framed}
$\:$
\end{framed}
\else
%\noindent
\begin{framed}
\noindent 
	{\bf 	From Machine Learning to Automated Machine Learning.}\\[1.5ex]
	The problem of supervised machine learning can be 
	formalized as finding a function $f$ that maps 
	possible input instances from a given set $X$ to possible target values from a set $Y$ such that a loss function is minimized on a given set of examples, i.e., as determining $\arg \min_{f\in F} L(f,E)$, where 
	$F$, referred to as the hypothesis space, is a set of 
	functions from $X$ to $Y$, $L$ is the loss function, and $E$ is the set of examples (or training data), comprised of input instances and target values.

	When $Y$ is a set of discrete values, this problem is called  {\em (supervised) classification};
	when it is the set of real numbers, it is known as {\em (supervised) regression}.  
	Popular loss functions include
	cross-entropy for classification and  mean squared error for regression.
	
	In this formulation, different hypothesis spaces $F$ can be chosen for a given supervised machine learning task. \sout{The functions in $F$ can be parametric 
	(where learning corresponds to finding values for a given set of parameters, such as the weights in logistic regression or neural networks) or non-parametric (e.g., decision trees or nearest neighbor procedures).} 
In addition to the parameters of a given model
(such as the connection weights in a neural network) that
determine a specific $f \in F$, there are typically further parameters that define
the function space $F$ (such as the structure of a neural network)
or affect the performance of the model induction process
(such as learning rates). 
	Generally, these \emph{hyperparameters} can be of different types 
	(such as real numbers, integers or categorical) and may be subject to complex dependencies
	(such as certain hyperparameters only being active when others take certain values). 
    Because the performance of modern machine learning techniques critically depends on hyperparameter settings, there is a growing need for hyperparameter optimization techniques.
    At the same time, because of the complex dependencies between hyperparameters,  sophisticated methods are needed for this optimization task.
	
   Human experts not only face the problem of determining performance-optimizing hyperparameter settings, but the choice of the class of machine learning models to be used in the first place, and the algorithm used to train these.  
In \emph{automated machine learning (AutoML)} all these tasks, often along with feature selection, ensembling and other operations closely related to model induction, are fully automated, such that performance is optimized for a given use case, e.g., in terms of the prediction accuracy achieved based on given training data.

\end{framed}
\fi

Faced with the choice from a large set of machine learning algorithms and an even larger space of hyperparameter settings, even seasoned experts often have to resort to experimentation to determine what works best in a given use case. Automated machine learning attempts to automate this process, and thereby not only spares experts the time and effort of extensive, often onerous experimentation, but also enables non-experts to obtain 
substantially better performance than otherwise possible. 
AutoML systems often achieve these advantages at rather high computational cost.

It is worth noting that AutoML falls squarely into the first form of automation, {\em mechanization},  as discussed in the introduction. At the same time, it can be seen as yet another level of abstraction over a series of automation stages. First, there is the well-known use of programming for automation. Second, machine learning automatically generates hypotheses and predictive models, which typically take the form of algorithms (e.g., in the case of a decision tree or a neural network); therefore, machine learning methods can be seen as meta-algorithms that automate programming tasks, and hence ``automate automation''.
And third, automated machine learning makes use of algorithms that select and configure machine learning algorithms---i.e., of meta-meta-algorithms that can be understood as automating the automation of automation.

AutoML systems have been gradually automating more and more of these tasks: model selection, hyperparameter optimization and feature selection. 
Many of these systems also deal with automatically selecting learning algorithms based on properties (so-called meta-features) of given data sets, building on the related area of  meta-learning \cite{brazdil2008metalearning}. In general, AutoML systems are based on sophisticated algorithm configuration methods, such as SMAC  (sequential model-based algorithm configuration) \cite{hutter2011sequential}, learning to rank and Monte-Carlo Tree Search  \cite{rakotoarison-ijcai2019-457}. 

So far, most work on AutoML has been focused on supervised learning.
Auto-WEKA \cite{ThoEtAl13}, one of the first AutoML systems, builds on the well-known Weka machine learning environment.
It encompasses all of the classification approaches implemented in Weka's
standard distribution, including a large number of base classifiers,
feature selection techniques, meta-methods that can build on any of
the base classifiers, and methods for constructing ensembles. 
Auto-WEKA 2 \cite{KotEtAl17} additionally deals with regression
procedures, and permits the optimization of any of the performance metrics supported by Weka 
through deep integration with the Weka environment.
The complex optimization process at the heart of Auto-WEKA is carried out by SMAC. 
Auto-sklearn \cite{FeuEtAl15} makes use of the Python-based machine learning toolkit scikit-learn, 
and is also powered by SMAC. 
Unlike Auto-WEKA, Auto-sklearn first determines multiple base learning procedures, which are then greedily combined into an ensemble.

These AutoML methods are now making their way into large-scale commercial applications enabling, for example, non-experts to more easily build  relatively complex supervised learning models. 
Recent work on AutoML includes neural architecture search (NAS), which automates key aspects of the design of neural network architectures, particularly (but not exclusively) in the area of deep learning (see, e.g., \cite{LiuEtAl18}). 
Google Cloud's proprietary AutoML tool, launched in early 2018, falls into this important, but restricted class of AutoML approaches.
Similarly, Amazon SageMaker, a commercial service launched in late 2017, provides some AutoML functionality
and covers a broad range of machine learning models and algorithms. 

The impressive performance levels reached by AutoML systems are evident in the results from recent competitions~\cite{GuyEtAl15}. 
Notably, Auto-sklearn significantly outperformed human experts in the human track of the 2015/2016 ChaLearn AutoML Challenge. 
Yet, results from the same competition suggest that human experts can achieve significant performance improvements by manually tweaking the classification and regression algorithms obtained from the best AutoML systems. 
Therefore, there appears to be considerable room for improvement in present AutoML systems for standard supervised learning settings.

Other systems, such as the Automatic Statistician \cite{lloyd2014a},
handle different kinds of learning problems, such as time series,
finding not only the best form of the model, but also its parameters.
We will revisit this work in the section on \emph{\Exploitation}; an
example of automatically fitting a time series is shown in
Figure \ref{fig:automaticstatistician}.

The automation of model building tasks in data science has 
been remarkably successful, especially in supervised learning.
We believe that the main reason for this lies in the fact that these tasks are usually very precisely specified and have relatively little dependence on the given domain (see also Figure~1), which renders them particularly suitable for 
{\em mechanization}.
Conversely, tasks beyond standard supervised learning, such as  unsupervised learning, 
have proven to be considerably harder to automate effectively, 
because the optimization goals 
are more subjective and domain-dependent, involving trade-offs between accuracy, efficiency, robustness, explainability, fairness, and more. 
Such machine learning methods, which are often used for feature engineering, domain understanding, data transformation, etc., thus extend into the remaining three quadrants, 
where we believe that more progress can be obtained using the other two kinds of automation seen in the introduction: {\em composition} and {\em assistance}.

\section*{Data Engineering: Big Gains, Big Challenges  \label{sec:dataeng}}

A large portion of the life of a data scientist is spent acquiring, organizing,
and preparing data for analysis, tasks  we collectively term \emph{data engineering}\footnote{Data wrangling
and data cleansing are terms that are also associated with many of
these stages.}. 
The goal of data engineering is to create consolidated data 
that can be used for further analysis or exploration. 
This work can be time-consuming and laborious, making it 
a natural target for automation.  
However, it faces the challenge of being more open-ended, as per its location in Figure \ref{fig:quadrant}.

To illustrate the variety of tasks involved in data engineering, 
consider the study \cite{bjorkman2018plant}
of how shrub growth in the
tundra has been affected by global warming.  
Growth is measured across a number of traits, 
such as plant height, leaf area, etc. 
To carry out this analysis, the authors had to: (i) integrate temperature
data from another dataset (using latitude, longitude and date
information as keys); (ii) standardize the plant names, which were
recorded with some variations (including typos); (iii) handle 
problems arising from being unable to integrate the 
temperature and biological data if key data was missing;
and (iv) handle anomalies by removing
observations of a given taxon that lay more than eight standard
deviations from the mean.

In general, there are many stages in the data engineering process,
with potential feedback loops between them. These can be divided into
three high-level themes, around (1) \emph{data organization},
(2) \emph{data quality} and (3) \emph{data
transformation} \cite{nazabal-williams-etal-20}, as we discuss in turn
below. For a somewhat different structuring of the relevant issues,
see, e.g., \cite{heer-hellerstein-kandel-19}.

Beginning with the first stage, \emph{data organization}, one of the first steps  is typically \emph{data parsing}, determining the structure of the data so that it can be imported into a  data analysis software environment or package.
Another common step is \emph{data integration}, which  aims to acquire, consolidate and restructure the data, 
which may exist in
heterogeneous sources (e.g., flat files, XML, JSON, relational databases), and in different locations. It may also require the alignment of data at different spatial resolutions or on different time-scales.  Sometimes
the raw data may be available in unstructured or semi-structured
form. In this case it is necessary to carry out \emph{information
extraction} to put the relevant pieces of information into tabular
form. For example, natural language processing can be used for
information extraction tasks from text 
(e.g., identifying names of people or places).  Ideally, a
dataset should be described by a data dictionary or \emph{metadata}
repository, which specifies information such as the meaning and type
of each attribute in a table. However, this is often missing or
out-of-date, and it is necessary to infer such information from the
data itself. For the data type of an attribute, this may be at the
syntactic level (e.g., the attribute is an integer or a calendar
date), or at a semantic level (e.g., the strings are all countries and
can be linked to a knowledge base, such as DBPedia), as in
\citep{chen-jimenez-ruiz-horrocks-sutton-19}.

FlashExtract \citep{FlashExtract} is an
example of a tool that provides assistance to the analyst for
the information extraction task. It can learn how to extract records from
a semi-structured dataset using a few examples; see Figure
\ref{fig:flashextract} for an illustration. A second assistive tool is
DataDiff \citep{sutton-hobson-geddes-caruana-18}, which 
integrates data that is received in installments, e.g., by means of monthly or annual
updates. It is not uncommon that the structure of the data may change
between installments, e.g., an attribute is added if new information is
available.  The challenge is then to integrate the new data by
matching attributes between the different updates.
DataDiff uses the idea that the
statistical \emph{distribution} of an attribute should remain similar
between installments \sout{is used} to automate the process of matching.

\ifdefined\NOFLOATS
\else
\begin{figure}[H]
\begin{center}
%\hspace{1cm}
\includegraphics[width=0.75\textwidth]{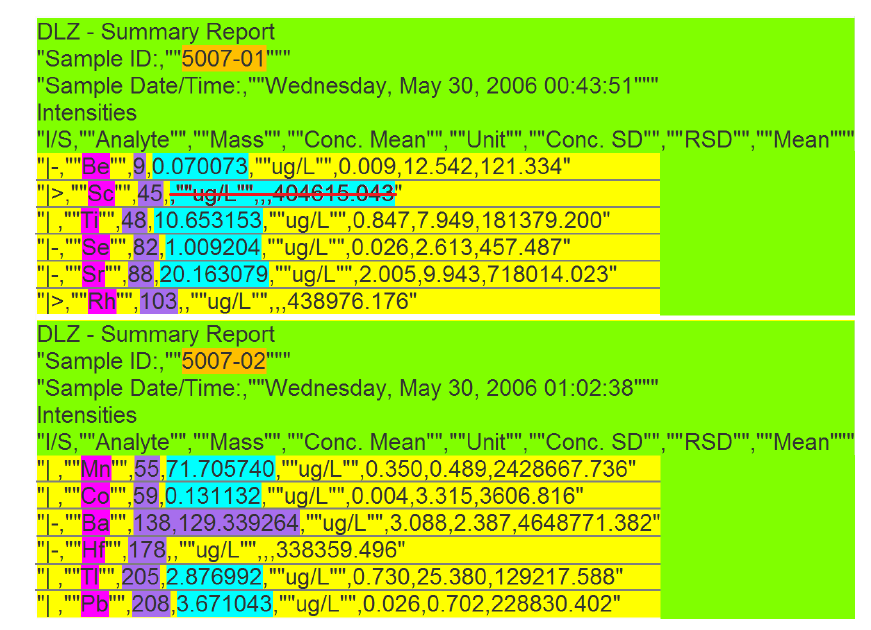} \hfill
\end{center}
 %\hspace{1cm}
 \vspace{-0.75cm}
\caption{FlashExtract (figure from \citep{FlashExtract}): 
After separating attributes by colours, FlashExtract can recognize examples (such as Be, 9 and 0.070073; and Ti, 48 and 10.653153) and counter-examples (such as the part struck through in red), in order to induce a program that is able to identify  other occurrences of these fields and put them in a spreadsheet or table for further processing.    }
\label{fig:flashextract}
\end{figure}
\fi

In the second stage of data engineering, \emph{data quality}, a common task is  \emph{standardization},
involving  processes that convert entities that have more than one
possible representation into a standard format. These might e.g., be
phone numbers with formats like ``(425)-706-7709'' or ``416 123 4567'', or text, e.g.,
``U.K.'' and ``United Kingdom''. In the latter case, standardization
would need to make use of ontologies that contain information about
abbreviations.  
\emph{Missing data} entries may be
denoted as ``NULL'' or ``N/A'', 
but could also be indicated by other
strings, such as ``?'' or ``-99''.  
This gives rise to two
problems: the identification of missing values, and handling
them downstream in the analysis.  
Similar issues of identification and repair arise if the data is corrupted by \emph{anomalies} or outliers. 
Because much can be done by looking at the distribution of the data only, many data science tools include (semi-)automated algorithms
for data imputation and outlier detection, which would fall under the {\em mechanization} or {\em assistance} forms of automation.

Finally, under the \emph{data transformation} heading,
we consider processes at the interface between 
data engineering and model building or data exploration. 
Feature engineering involves the construction of features based on the analyst's knowledge or beliefs.
When the data involves sensor readings, images or other
low-level information, signal processing and computer vision techniques may be required to determine or create meaningful features that can
be used downstream. 
Data transformation also includes instance selection, e.g., 
for handling imbalanced data or addressing unfairness due to bias.

As well as the individual tasks in data engineering, where we have seen
that \emph{assistive} automation can be helpful, there is also the
need for the \emph{composition} of tasks. 
Such a focus on composition is found, for example, in Extraction,
Transformation and Load (ETL) systems, which are usually supported by
a collection of scripts that combine data scraping, source
integration, cleansing and a variety of other transformations on the data. 

An example of a more integrated approach to data engineering, which shows aspects of both
compositional and assistive automation, is the \emph{predictive interaction} framework discussed 
in \citep{heer-hellerstein-kandel-15}. 
This approach provides
interactive recommendations to the analyst 
about which data engineering operations to apply at a
particular stage, in terms of an appropriate domain specific language, ideas that form the basis of the commercial data wrangling software from Trifacta. 
Another interesting direction is based on a concept known as \emph{data programming}, which exploits domain knowledge by means of programmatic creation and modeling of data sets for supervised machine learning tasks \cite{RatEtAl16}.

Methods from AutoML could potentially also help with data engineering. For instance, Auto-sklearn \cite{FeuEtAl15}  includes several pre-processing steps in its search space, such as simple missing data
imputation and one-hot encoding of categorical features. However, these steps can be seen as small parts of the data quality theme, which
can only be addressed once the many issues around data organization and other data quality steps (e.g., the \emph{identification} of missing data) have been carried out. 
These earlier steps are more open-ended and thus much less amenable to inclusion in the AutoML search process.

While many activities related to  storage, aggregation and data cleaning have been significantly automated by recent database technology, significant challenges remain, due to the fact that data engineering is often an iterative process over representation and integration steps, involving data from very different sources and in different formats, with feedback loops between
the steps that trigger new questions 
(see, e.g., \cite{heer-hellerstein-kandel-19}).  
For instance, in the Tundra example, one must know (i) that it is important to integrate the biological and temperature data, (ii) that the data must already be in a close-enough format for the transformations to apply, and (iii) that domain knowledge is needed to fuse variant plant names. 

As all these data engineering challenges occupy large amounts of
analyst time, there is an incentive to automate them as much as
possible, as the gains could be high. However, doing this poorly can
have a serious negative impact on the outcome of a data science
project.  We believe that many aspects of data engineering are
unlikely to be fully automated in the near future, except for a few
specific tasks, but that further developments in the direction of both
assistive and compositional semi-automation will nonetheless be
fruitful.

\section*{Data Exploration: More Assistance Than Automation \label{sec:dataunder}}

Continuing our discussion of the quadrants in
Figure \ref{fig:quadrant}, we next focus on data exploration.  The
purpose of data exploration is to derive insight or make discoveries
from given data (e.g., in a genetics domain, {\em understanding} the
relation between particular genes, biological processes, and
phenotypes), often to determine a more precise goal for a subsequent
analysis (e.g., in a retailing domain, discovering that a few
variables explain why customers behave differently, suggesting a
segmentation over these variables).  This key role of human insight in
data exploration suggests that the form of automation that prevails in
this quadrant is {\em assistance}, by generating elements that can
help humans reach this insight.  We will collectively refer to all
these elements that ease human insight as {\em patterns}, capturing
particular aspects or parts of the data that are potentially striking,
interesting, valuable, or remarkable for the data analyst or domain
expert, and thus worthy of further investigation or exploitation.
Patterns can take many forms, from the very simple (e.g., merely
reporting summary statistics for the data or subsets thereof), to more
sophisticated ones (e.g., communities in networks or low-dimensional
representations).

The origins of contemporary data exploration techniques can be traced back to Tukey and Wilk~\cite{tukey1966data}, who stressed the importance of human involvement in data analysis generally speaking, and particularly in data analysis tasks aiming at `exposing the unanticipated'---later coined Exploratory Data Analysis (EDA) by Tukey \cite{tukey1977eda} and others.

The goal of EDA was described as hypothesis generation, and was contrasted with confirmatory analysis methods, such as hypothesis testing, 
which would follow in a second step.
Since the early days of EDA in the 1970s, 
the array of methods for data exploration, the size and complexity of data, and the available memory and computing power  have all vastly increased.
While this has created unprecedented new potential, it comes at the price of greater complexity, thus creating a need for automation to assist the human analyst in this process.

As an example, 
the `Queriosity' system \citep{wasay2015queriosity} provides a vision of automated data exploration as a dynamic and interactive process, allowing the 
system to learn to understand
the analyst's evolving background and intent, in order to enable it
to proactively show `interesting' patterns. 
The FORSIED framework \citep{de2013subjective} has a similar goal, formalizing the data exploration process as an interactive exchange of information between data and data analyst, accounting for the analyst's prior belief state.
These approaches stand  in contrast to the 
more traditional approach to data exploration, where the analyst  repeatedly queries the data for specific patterns in a time- and
labor-intensive process, in the hope that some of the patterns turn out to be
interesting. This vision means that the automation of data exploration requires the identification of what the analyst knows (and does not know) about the domain, so that knowledge and goals, and not only patterns, can be articulated by the system. 

To investigate the extent to which automation is possible and
desirable, without being exhaustive, it is helpful to identify five
important and common subtasks in data exploration, as illustrated for
a specific use case (social network analysis) in the associated box.
These five problems are discussed at a generic level below.

$\:$

\ifdefined\NOFLOATS
\framebox{
$\:$
}
\else
\noindent
\framebox{\parbox{\textwidth}{
{\bf Five data exploration subtasks in social network analysis.}\\
Computational social scientists may wish to explore a social network
to gain an understanding of the social interactions it describes. 
For example, an analyst may decide to look for community patterns, formalized as subsets of the nodes and the edges connecting them.
 In the broad context of data exploration, five subtasks that can potentially be automated are outlined as follows:
\begin{enumerate}
    \item {\it Form of the pattern.} Options include the network's high-level topology, degree distribution, clustering coefficient, or the existence of dense subnetworks (communities) as considered here by way of example.  
    \vspace{-0.1cm}
    \item {\it Measuring pattern `interestingness'.} Interestingness can be quantified as the number of edges or the average node degree within the community, the local modularity, or subjective measures that depend on the analyst's prior knowledge, 
    or measures developed from scratch.
    \vspace{-0.1cm}
    \item {\it Algorithmic strategy.} Optimizing the chosen measure can require numerical linear algebra, graph theory, heuristic search (e.g.,\ beam search), or bespoke approaches.
    \vspace{-0.1cm}
    \item {\it Pattern presentation.} The most interesting communities can be presented to the analyst as lists of nodes, by marking them on a suitably permuted adjacency matrix, or using other visualizations of the network.
    \vspace{-0.1cm}
    \item {\it Interaction.} Almost invariably, the analyst will want to iterate on some of the subtasks, e.g. to retrieve more communities, or to explore other pattern forms.
\end{enumerate}
}}
\fi

$\:$

The \emph{form of the patterns} (subtask 1) is often dictated by the data analyst, i.e., user involvement is inevitable in choosing this form. Indeed, certain types of pattern may be more intelligible to the data analyst, or may correspond to a model of physical reality. As illustrated in the box, a computational social scientist may be interested in finding dense subnetworks in a social network as evidence of a tight social structure.

There are often too many possible patterns. 
Thus, a \emph{measure to quantify how interesting any given set of patterns of this type is to the data analyst} is required (subtask 2). Here, `interestingness' could be defined in terms of coverage, novelty, reliability, peculiarity, diversity, surprisingness, utility, or actionability; moreover, each of these criteria can be quantified either objectively (dependent on the data only), subjectively (dependent also on the data analyst), or based on the semantics of the data (thus also dependent on the data domain) \cite{geng2006interestingness}. 
Designing this measure well is crucial but also highly non-trivial, making this a prime target for automation.
Automating this subtask may require understanding the data analyst's intentions or preferences \cite{ruotsalo2014interactive},
the perceived complexity of the patterns, and the data analyst's background knowledge about the data domain---all of which require interaction with the data analyst.
The latter is particularly relevant for the formalization of novelty and surprisingness in a subjective manner, and recent years have seen significant progress along this direction using information-theoretic approaches \cite{de2013subjective}.

The next stage (subtask 3) is to identify the \emph{algorithms needed to optimize the chosen measure.} 
In principle, it would be attractive to facilitate this task using higher-level automation, as done in AutoML.
However, considering the diversity of data across applications, the diversity of pattern types, and the large number of different ways of quantifying how interesting any given pattern is, there is a risk that different data exploration tasks may require different algorithmic approaches for finding the most interesting patterns. 
Given the challenges in designing such algorithms, we believe that more generic techniques or declarative approaches (such as inductive databases and probabilistic programming, covered in the final section of the paper) may be required to make progress in the {\em composition} and {\em assistance} forms of automation for this subtask.

The user interface of a data exploration system often \emph{presents the data, and identifies patterns within it, in a visual manner} to the analyst (subtask 4).
This makes it possible to leverage the strong perceptual abilities of the human visual system, as has been exploited and enhanced by decades of research in the visual analytics community \cite{keim2008visual}. 
At the same time, the multiple comparisons problem inherent in visual analysis may necessitate steps to avoid false discoveries~\cite{zgraggen2018investigating}.
Automating subtask 4 beyond some predefined visualisations (as in the 
Automatic Statistician, see Figure \ref{fig:automaticstatistician}) requires a good understanding of the particular perception and cognition capacities and preferences of each user, a question that 
also features prominently in the related area of explainable artificial intelligence, which we will discuss in the following section.  

Such visualizations and other kinds of tools for navigating the data must allow for \emph{rich and intuitive forms of interaction} (subtask 5), to mitigate the open-endedness of typical data exploration tasks. They must allow the analyst to follow leads, verify or refine hypotheses by drilling deeper, and provide feedback to the data exploration system about what is interesting and what is not. 
A huge challenge for automation is how a novice data analyst could be given hints and recommendations of the type an expert might use, {\em assisting} in the process of data navigation, from the combinatorial explosion of ways of looking into the data and possible kinds of patterns. 
For instance, 
the SeeDB~\cite{vartak2015seedb} and Voyager~\cite{wongsuphasawat2015voyager}
systems interactively recommend visualizations that may be particularly effective, and \emph{Interactive intent modeling}~\cite{ruotsalo2014interactive} has been proposed to improve information-seeking efficiency in information retrieval applications.

Each of the five subtasks is challenging on its own and 
contains many
design choices that may require expert knowledge. 
We argue that the limitations of current AI techniques in acquiring and dealing with human knowledge in real-world domains are the main reason why automation in this quadrant 
is typically in the form of {\em assistance}. 
Meanwhile, we should recognize that the above
subtasks are not independent, as they must combine, through the {\em composition} form of automation, to effectively assist the
data analyst, and non-expert users, in their search for new insights and discoveries.

\section*{Exploitation: Automation within the Real World \label{sec:inter-deploy}}

The bottom right \quadrant in Figure \ref{fig:quadrant} is usually reached when the insights from other tasks have to be translated back to the application domain, often -- but not always -- in the form of predictions or, more generally, decisions. 
This \quadrant deals with extracted \emph{knowledge} and less with data, 
involving the understanding of the patterns and models, 
publishing them as building blocks for new discoveries (e.g.,  in scientific papers or reports), 
putting them into operation, 
validating and monitoring their operation, and ultimately revising them. 
This quadrant 
is usually less open-ended, so it is no surprise that some specific activities here, such as reporting and maintenance, can be automated to a high degree.

The interpretation of the extracted knowledge is closely related to the area of explainable or interpretable machine learning. Recent surveys cover different ways in which explanations can be made, but 
do not analyze the degree and form of automation (see, e.g., \cite{guidotti2018survey}). 
Clearly, the potential for automation depends strongly on whether a generic explanation of a model (global explanation) or a single prediction (local explanation) is required, and  whether the explanation has to be customized for or interact with a given user, by adaptation to their background, expectations, interests and personality. 
Explanations must go beyond the inspection or transformation of models and predictions, and should include the relevant variables for these predictions, the distribution of errors and the kind of data for which it is more or less reliable, the vulnerabilities of a model, how unfair it is, etc. 
A prominent example following the {\em mechanization} form of automation is the Automatic Statistician\footnote{ \url{https://www.automaticstatistician.com/}} \cite{lloyd2014a}, 
which is able to produce a textual report on the model produced (for a
limited set of problem classes). Figure
\ref{fig:automaticstatistician} shows a fragment of such a report,
including graphical representations and textual explanations of the
most relevant features of the obtained model and its behavior.

\ifdefined\NOFLOATS
\else
\begin{figure}[t]
\begin{center}
\fbox{\includegraphics[width=0.73\textwidth]{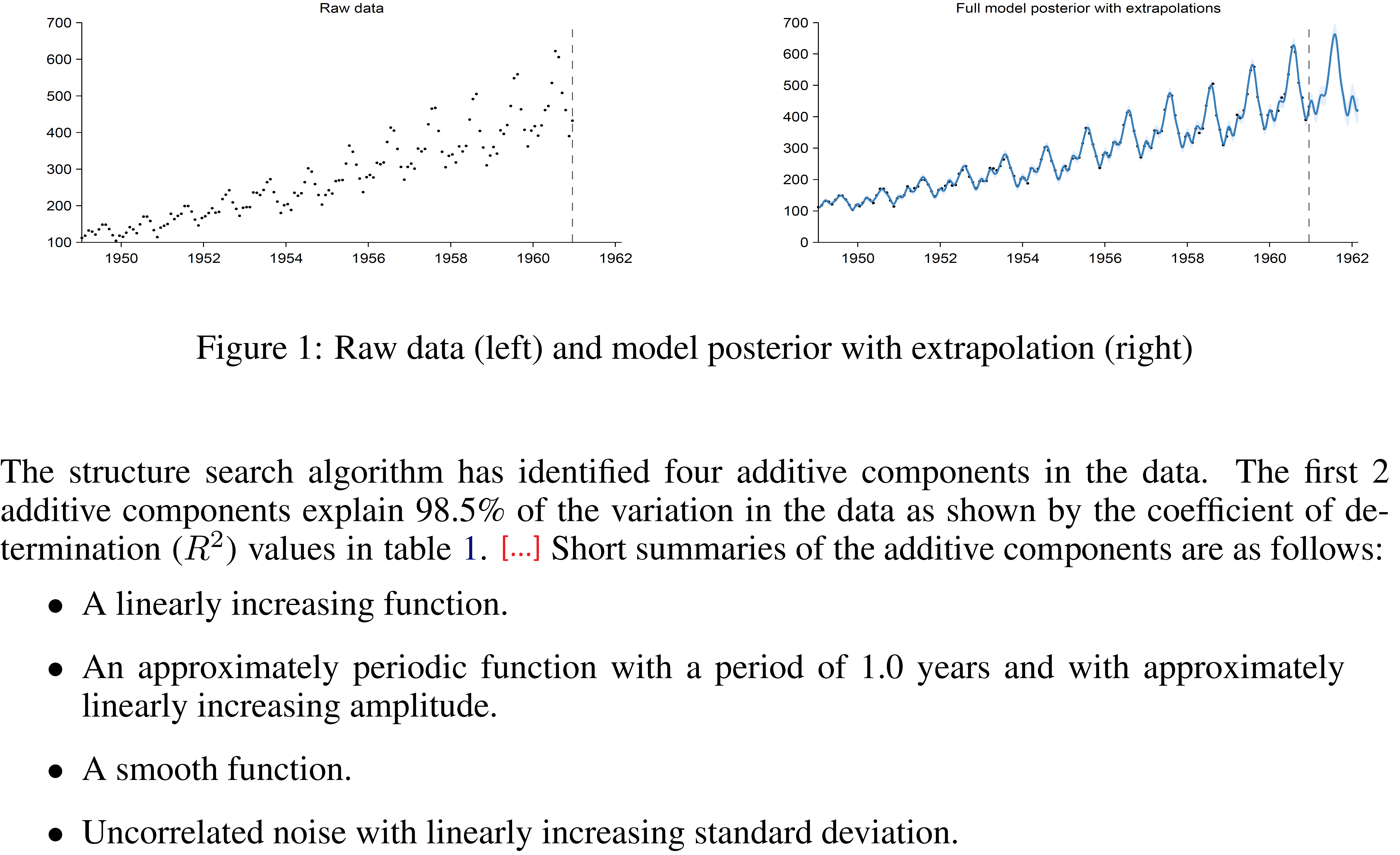}} \hfill
\end{center}
 \vspace{-0.7cm}
\caption{A fragment of the Automatic Statistician report for the
  ``airline'' dataset, which considers airline passenger volume over
  the period from 1949 to 1961 (from \cite{lloyd2014a}).}
\label{fig:automaticstatistician}
\end{figure}
\fi

We believe that fully understanding the behavior and effect of the models and insight produced in earlier stages of the data science pipeline 
is an integral part of the validation of the entire process, and key to a successful deployment. However, `internal'  evaluation, which is usually coupled with model building or carried out immediately after, is done {\em in the lab}, trying to maximize some metric on held-out data. 
In contrast, validation {\em in the real world} refers to meeting some goals, with which the data, objective functions and other elements of the 
process may not be perfectly aligned. Consequently, this broad perspective of the `external' validation poses additional challenges for automation, as domain context plays a more important role (Figure \ref{fig:quadrant}). 
This is especially the case in areas where optimizing for trade-offs between accuracy and fairness metrics may still end up producing undesirable global effects in the long term, or areas such as 
safety-critical domains, where experimenting with the actual systems is expensive and potentially dangerous, e.g., in medical applications or autonomous driving. 
A very promising approach to overcome some of these challenges is the use of simulation, where an important part of the application domain is modeled, be it a hospital \cite{elbattah2018analytics} or a city. The concept of `digital twins' \cite{tao2019make} 
allows data scientists to deploy their models and insights in a
digital copy of the real world, to understand and exploit causal
relations, and to anticipate effects and risks, as well as to optimize
for the best solutions. Optimization tools that have proven so useful
in the AutoML scenario can be used to derive globally optimal
decisions that
translate from the digital twin to the real world, provided the simulator is an accurate model at the required level of abstraction. The digital twin can also be a source of simulated data for further iterations of the entire data science process. 

Deployment becomes more complex as more decisions are made, models are produced and combined, and many users are involved. Accordingly, we contend that automating model \emph{maintenance and monitoring} is becoming increasingly relevant. This includes tracing all the dependencies between models, insights and decisions that were generated during training and operation, especially if re-training is needed \cite{sculley2015hidden}, 
resembling software maintenance in several ways. 
Some aspects of monitoring trained models seem relatively straightforward and automatable, 
by re-evaluating indicators (metrics of error, fairness, etc.) periodically and flagging important deviations, as a clear example of the {\em assistive} form of automation, 
which allows for extensive reuse. 
Once models are considered unfit or degraded, retraining to some new data that has shifted from the original data seems easily mechanizable (repeating the experiment), but it depends on whether the operating conditions that were used initially still hold after the data shift. 
Reliable and well-understood models can often be reused even in new or changing circumstances, through domain adaptation, transfer learning, lifelong learning, or reframing \cite{hernandez2016reframing}; 
this represents a more compositional form of automation. 

Data science creates a large amount of patterns, models, decisions and
meta-knowledge. 
The organization and reuse of models and patterns can be automated to some degree 
via 
inductive databases (which we discuss in the next section), 
via  specialized \emph{databases of models} (e.g., machine learning model management \cite{vartak2016m}), 
or by means of large-scale experimentation platforms, such as OpenML\footnote{\url{www.openml.org} \cite{vanschoren2014openml}}. In the end, we believe that the automation of {\em knowledge} management and analysis for and from data science activities will be a natural evolution of the automation of {\em data} management and analysis.

\section*{Perspectives and Outlook 
\label{sec:chal-outlook}}

The quest for automation, in the broad context of data analysis and
scientific discovery, is not new, spanning 
decades of 
work in statistics, artificial intelligence (AI), 
databases, and programming languages. We now visit each of these
perspectives in turn, before drawing some final conclusions.

First, there is a long tradition in AI of attempts to automate the
scientific discovery process. Many researchers have tried to
understand, model and support a wide range of scientific processes
with AI, including approaches to leverage cognitive models for
scientific discovery (such as Kepler's laws)
\cite{langley1987scientific}. 
More recent and operational models of scientific discovery include robot
scientists \cite{King}, which are robotic systems that design and
carry out experiments in order to find models or theories, e.g., in the life sciences. While these attempts included experimental design and not only  observational data, they were specialized to particular domains, reducing the challenges of the domain context (the vertical dimension in Figure \ref{fig:quadrant}). 
Many important challenges remain in this area, including the induction or revision of theories or models from very sparse data; the transfer of knowledge between domains (which is known to play an important role in the scientific process); the interplay between the design of methodology, including experiments, and the induction of knowledge from data; and the interaction between scientists and advanced computational methods designed to support them in the scientific discovery process.

Second, there were efforts in the 1980s and 1990s 
at the interface of statistics and AI to develop software systems that 
would build models or explore data, often in an interactive manner, using heuristic search or planning based on expert knowledge 
(e.g.,
\cite{gale1987statistical,st1998intelligent}). This line of research
ran up against the limits of knowledge representation, 
which proved inadequate to
capture 
the subtleties of the statistical strategies used by expert data analysts. 
Today, the idea of a 
`mechanized' statistical data analyst is still
being pursued (see, e.g., the Automatic Statistician \cite{lloyd2014a}),
but with the realization that statistical modeling often relies
heavily on human judgement in a manner that is not easy to capture
formally, beyond the top right quadrant in Figure \ref{fig:quadrant}.  
It is then the {\em composition} and {\em assistance} forms of automation that are still targeted when 
modular data analytic operations are combined into plans or workflows 
in
current data science platforms, such as KNIME and Weka, or in the form of  
intelligent data science assistants
\cite{serban-vanschoren-kietz-bernstein-13}.

Third, in a database context, the concept of inductive query languages allows a user to query the models and patterns that are held in the data. 
Patterns and models become ``first-class citizens'' with the hope of reducing many activities in data science to a querying process, in which the insights obtained from one query lead to the next query, until the desired patterns and models have been found. These systems are typically based on  extensions of SQL and other  relational database languages (see, e.g.,  \cite{blockeel2012inductive}). 
Doing data science as querying or programming may help bridge the composition and mechanisation forms of automation.  

Fourth, in recent years, there 
has been an increasing attention on probabilistic programming
languages, 
which allow the expression and
learning of complex probabilistic models, extended or combined with first-order logic    
\cite{raedt2016statistical}.
Probabilistic programming languages have been used inside tools for
democratizing data science, such as BayesDB
\cite{mansinghka2015bayesdb}  
and Tabular \cite{gordon2014tabular},
which build probabilistic models on top of tabular databases and
spreadsheets. Probabilistic programming can also, for example, propagate uncertainty from an imputation method for missing data into the predictive analysis and incorporate background knowledge into the analysis. This may support a more holistic view of automation by increasing the integration of the four quadrants in Figure \ref{fig:quadrant}, which may mutate accordingly.

All four of these approaches have had some success in specific domains or standard situations, but still lack the generality and flexibility needed for broader applications in data science, as the discipline incorporates new methods and techniques at a pace that these systems cannot absorb. More scientific and community developments are needed to bridge the  
gap between how data scientists actually conduct their work and the
level of automated support that such approaches can
provide. Table~\ref{tab:challenges} presents a series of
indicative technical challenges for automating data science.

\begin{table}[!h]
    \centering
    \begin{footnotesize}
    \begin{tabular}{|p{2.45cm}|p{6.7cm}|p{1.31cm}|p{1.31cm}|p{1.31cm}|}
     \hline
        {{\bf Quadrant}} & {{\bf Challenge}} & {{\bf Mech\-a\-ni\-za\-tion}} &
        {{\bf Com\-po\-si\-tion}} & {{\bf As\-sis\-tance}}    \\    \hline %        \\[-1ex]
        {Generic} & {Enhancing human-AI collaboration, by incorporating domain context for interactively defining and refining the goal of data science activities.} & {} & {$\quad\;\;\times$} & {$\quad\;\;\times$}    \\ \hline
        {Generic} & {Addressing ethical, privacy \& legal issues in the automation of data science.} & {$\quad\;\;\times$} & {$\quad\;\;\times$} & {$\quad\;\;\times$}    \\ \hline
        {Model building} & {Extending AutoML to tasks beyond supervised learning.} & {$\quad\;\;\times$} & {} & {}    \\ \hline
        {Model building / Data engineering} & {Generating meaningful features, taking into account domain context and task.} & {$\quad\;\;\times$} & {} & {$\quad\;\;\times$}    \\ \hline 
        {Data engineering} & {Streamlining the ETL (Extract, Transform, Load) process by using pipeline schemas and reusing preprocessing subcomponents} & {} & {$\quad\;\;\times$} & {}    \\ \hline
        {Data engineering} & {Expediting the data cleaning, outlier detection and data imputation processes.} & {$\quad\;\;\times$} & {$\quad\;\;\times$} & {$\quad\;\;\times$}    \\ \hline
        {Data exploration} & {Supporting the design of interactive data and pattern visualizations.} & {} & {} & {$\quad\;\;\times$}    \\ \hline
        {Data exploration} & {Developing human-AI collaborative systems for data and pattern exploration.} & {} & {$\quad\;\;\times$} & {$\quad\;\;\times$}    \\ \hline
        {Exploitation} & {Generating collaborative reports and presentations, facilitating the interrogation, validation and explanation of models and results.} & {$\quad\;\;\times$} & {} & {$\quad\;\;\times$}    \\ \hline
        {Exploitation} & {Dealing with concept drift, monitoring the interaction of several data science models, and assessing their effects more globally.} & {$\quad\;\;\times$} & {} & {$\quad\;\;\times$}    \\ 
        \hline
    \end{tabular}
    \end{footnotesize}
    \caption{
    Selected research challenges
in automating data science, with their associated quadrants and likely forms of automation (mechanization, composition and assistance).}
    \label{tab:challenges}
\end{table}

While AutoML will continue to be a flagship example for automation in data science, we expect most progress 
in the following years to involve stages and tasks other than modeling. 
Capturing information about how data scientists work, and how data science projects evolve from conception to deployment and maintenance, will be key for more ambitious tools. Progress in areas of AI such as reinforcement learning can accelerate this.

It is important to raise awareness of the potential pitfalls and side effects of higher levels of automation in data science. These include over-reliance on the results obtained from systems and tools; the introduction of errors that are subtle and difficult to detect; and cognitive bias towards certain types of observations, models and insights facilitated by existing tools. Also, data science tools in the context of human-AI collaboration are seen as displacing the work practice of data scientists, leading to new roles \cite{wang2019human}. Similarly, this collaborative view suggests new forms of interaction between data scientists and machines, as these become proactive assistants rather than tools \cite{amershi2019guidelines}.

With all of this in mind, we cautiously make the following
predictions. First, it seems likely that there will continue to be
useful and significant advances in the automation of data science in the three most accessible  quadrants in Figure 1: data engineering (e.g., automation of inference about
missing data and of feature construction), model building (e.g., automated
selection, configuration and tuning beyond the current scope of AutoML),
and exploitation (e.g., automated techniques for model diagnosis and summarization).   
Second, for the most challenging quadrant of data exploration, and for tasks in the other quadrants where representation of  domain knowledge and goals is needed, 
we anticipate that progress will require more effort. And third,
across the full spectrum of data science activities, we see great
potential for the assistance form of automation, through systems that
complement human experts, tracking and analyzing workflows, spotting
errors, detecting and exposing bias, and providing high-level advice.
Overall, we expect an
increasing demand 
for methods and tools that are better integrated with  human experience and domain expertise, with an emphasis on complementing and enhancing the work of human experts rather than on full mechanization.

\ifdefined\NOREFERENCES
%\section*{References}
\else
\def\bibfont{\footnotesize}
\bibliographystyle{plainnat}  % If we use natbib for \citeauthor, etc., we need plainnat instead of plain.
{\footnotesize 
%{\small 
\bibliography{ms}
}
%endgroup

$\:$

\noindent
\hrulefill

\section*{Authors}

{\bf Tijl De Bie} (\url{tijl.debie@ugent.be}) is full professor in the Internet and Data Lab (IDLab) at Ghent University.

\noindent{{\bf Luc De Raedt} (\url{luc.deraedt@kuleuven.be}) is full professor at the Department of Computer Science and Director of the KU Leuven Institute for AI at KU Leuven, Belgium, and Wallenberg Guest Professor at \" Orebro University in Sweden}

\noindent{\bf Jos\'e Hern\'andez-Orallo} (\url{jorallo@upv.es}) is full professor at the Valencian Research Institute for Artificial Intelligence, Universitat Polit\`ecnica de Val\`encia, Spain.

\noindent{\bf Holger H. Hoos} (\url{hh@liacs.nl}) is Professor of Machine Learning
at the Leiden Institute of Advanced Computer Science (LIACS) at Leiden University, The Netherlands, and Adjunct Professor of Computer Science at the University of British Columbia in Vancouver (BC), Canada.

\noindent{\bf Padhraic Smyth} (\url{smyth@ics.uci.edu}) is Chancellor's Professor in the Computer Science and Statistics Departments at the University of California, Irvine, USA.

\noindent{\bf Christopher K. I. Williams} (\url{ckiw@inf.ed.ac.uk}) is
Professor of Machine Learning in the School of Informatics, University
of Edinburgh, United Kingdom, and a Turing Fellow at the Alan Turing
Institute, London, UK.

\vspace{0.5cm}
\noindent {\em The authors thank the anonymous referees for their comments, which helped to improve the paper.}

\section*{Funders}

\begin{small}
\noindent
TDB: The European Research Council under the European Union's Seventh Framework Programme (FP7/2007-2013) / ERC Grant Agreement no. 615517. The Flemish Government under the ``Onderzoeksprogramma Artificiële Intelligentie (AI) Vlaanderen'' programme. The Fund for Scientific Research -- Flanders (FWO--Vlaanderen), project no. G091017N, G0F9816N, 3G042220.\\
\noindent
LDR : The research reported in this work was  supported by the European Research Council (ERC) under the European Union’s Horizon 2020 research and innovation programme (grant agreement No [694980] SYNTH: Synthesising Inductive Data Models), the EU H2020 ICT48 project ``TAILOR'', under contract \#952215; the Flemish Government under the ``Onderzoeksprogramma Artificiële Intelligentie (AI) Vlaanderen'' programme and the Wallenberg AI,
Autonomous Systems and Software Program (WASP) funded
by the Knut and Alice Wallenberg Foundation.\\
\noindent
JHO: Funder: EU (FEDER) and the Spanish MINECO, Grant: RTI2018-094403-B-C3.
Funder: Generalitat Valenciana, Grant:  PROMETEO/2019/098. Funder: FLI, Grant RFP2-152. Funder: MIT-Spain INDITEX Sustainability Seed Fund. Grant: FC200944. Funder: EU H2020. Grant:  ICT48 project ``TAILOR'', under contract \#952215.

\noindent
HHH: The research reported in this work was partially supported by the EU H2020 ICT48 project ``TAILOR'', under contract \#952215; by the EU project H2020-FETFLAG-2018-01, ``HumanE AI'', under contract \#820437, and by start-up funding from Leiden University.

\noindent
PS: This material is based on work supported by the US National Science Foundation under awards DGE-1633631, IIS-1900644, IIS-1927245, DMS-1839336, CNS-1927541, CNS-1730158,  DUE-1535300; by the US National Institutes of Health under award 1U01TR001801-01; by NASA under award NNX15AQ06A.

\noindent
CKIW: This work is supported in part by  grant EP/N510129/1 from the UK Engineering and Physical Sciences Research Council (EPSRC) to the Alan Turing Institute. He thanks the Artificial Intelligence for Data Analytics team at the Turing for many helpful conversations. 
\end{small}
\fi

\end{document}